\documentclass[twocolumn,superscriptaddress,aps,prb]{revtex4-1}

\usepackage{graphicx}
\usepackage[usenames,dvipsnames]{color}
\usepackage[normalem]{ulem}
\usepackage{amsmath}
\usepackage{multirow}

\begin{document}

\newcommand{\mnps}{MnPS$_3$}
\newcommand{\feps}{FePS$_3$}
\newcommand{\ceme}{cm$^{-1}$}
\newcommand{\mub}{$\mu_\mathrm{B}$}
\newcommand{\etal}{\textit{et al.}}
\newcommand{\blue}{\textcolor{blue}}

\title{Magnetoelastic interaction in the two-dimensional magnetic
  material MnPS$_3$ studied by first principles calculations and Raman
  experiments}

\author{Diana Vaclavkova}
\affiliation{Laboratoire National des Champs Magn\'etiques Intenses,
  CNRS-UGA-UPS-INSA-EMFL, Grenoble, France}

\author{Alex Delhomme}
\affiliation{Laboratoire National des Champs Magn\'etiques Intenses,
  CNRS-UGA-UPS-INSA-EMFL, Grenoble, France}

\author{Cl\'ement Faugeras} 
\affiliation{Laboratoire National des Champs Magn\'etiques Intenses,
  CNRS-UGA-UPS-INSA-EMFL, Grenoble, France}

\author{Marek Potemski}
\affiliation{Laboratoire National des Champs Magn\'etiques Intenses,
  CNRS-UGA-UPS-INSA-EMFL, Grenoble, France}
\affiliation{Institute of Experimental Physics, Faculty of Physics,
  University of Warsaw, 02-093 Warsaw, Poland}

\author{Aleksander Bogucki}
\affiliation{Institute of Experimental Physics, Faculty of Physics,
  University of Warsaw, 02-093 Warsaw, Poland}

\author{Jan Suffczy\'nski}
\affiliation{Institute of Experimental Physics, Faculty of Physics,
  University of Warsaw, 02-093 Warsaw, Poland}

\author{Piotr Kossacki}
\affiliation{Institute of Experimental Physics, Faculty of Physics,
  University of Warsaw, 02-093 Warsaw, Poland}

\author{Andrew R. Wildes}
\affiliation{Institut Laue-Langevin, Grenoble, France}

\author{Beno\^it Gr\'emaud}
\affiliation{Aix Marseille Univerist\'e, Universit\'e de
  Toulon, CNRS, CPT, Marseille, France}
\affiliation{MajuLab, CNRS-UCA-SU-NUS-NTU
  International Joint Research Unit, 117542 Singapore}
\affiliation{Centre for Quantum Technologies, National University of
  Singapore, 2 Science Drive 3, 117542 Singapore}

\author{Andr\'es Sa\'ul}
\affiliation{Aix-Marseille Universit\'e, Centre Interdisciplinaire de
Nanoscience de Marseille-CNRS (UMR 7325), Marseille, France}
\email{saul@cinam.univ-mrs.fr}

\begin{abstract}
  We report experimental and theoretical studies on the magnetoelastic
  interactions in \mnps.  Raman scattering response measured as a
  function of temperature shows a blue shift of the Raman active modes
  at 120.2 and 155.1 \ceme, when the temperature is raised across the
  antiferromagnetic-paramagnetic transition. Density functional theory
  (DFT) calculations have been performed to estimate the effective
  exchange interactions and calculate the Raman active phonon
  modes. The calculations lead to the conclusion that the peculiar
  behavior with temperature of the two low energy phonon modes can be
  explained by the symmetry of their corresponding normal coordinates
  which involve the virtual modification of the super-exchange angles
  associated with the leading antiferromagnetic (AFM) interactions.
\end{abstract}

\maketitle

\section{Introduction}
\label{sec:intro}

The research dedicated to the so called two-dimensional (2D) materials
\cite{Geim2013}
%
%
has been lately expanded to studies of a large class of layered
magnetic materials and their ultrathin crystals \cite{Gibertini2019}.
%
%
This, to great extent, is stimulated by the search for novel effects
in the magnetic ordering in the 2D limit and possible design of new
devices which take advantage of the spin degree of freedom. Relevant
research efforts are, in particular, being focused on
antiferromagnetic transition-metal thiophosphates
\cite{Mayorga-Martinez2017a},
%
%
with their representative example, the \mnps\ compound.
Changes in the magnetic ordering are often traced with Raman
scattering, a technique which appears to be especially well suited to
study small size and ultrathin samples. The elementary excitations
which are most often probed with Raman scattering are phonon
excitations. Answering the question how and which phonon modes appear
to be sensitive to the magnetic order is central for drawing
conclusions on the applicability of Raman scattering methods to study
magnetic phase transitions.

In this paper we present a first-principle study of magnetoelastic
interaction in \mnps\ antiferromagnets and compare our theoretical
expectations with the results of temperature dependent Raman
scattering measurements performed on these crystals. The aim of our
work is to identify the phonon modes which are sensitive to magnetic
order and to enlighten the mechanism of the phonon-spin coupling that
is at the origin of this sensitivity.

Experimentally, we confirm \cite{Materials2019} a robust sensitivity
of the Raman active phonon mode at 155.1 \ceme\ (at low temperatures)
to the antiferromagnetic to paramagnetic transition in \mnps\ and
clarify a similar property for the other, but weaker, phonon mode
observed at 120.2 \ceme. These experimental observations are well
accounted for by our theoretical calculations.

\begin{figure*}[htb!]
\begin{center}
  \includegraphics[width=\textwidth]{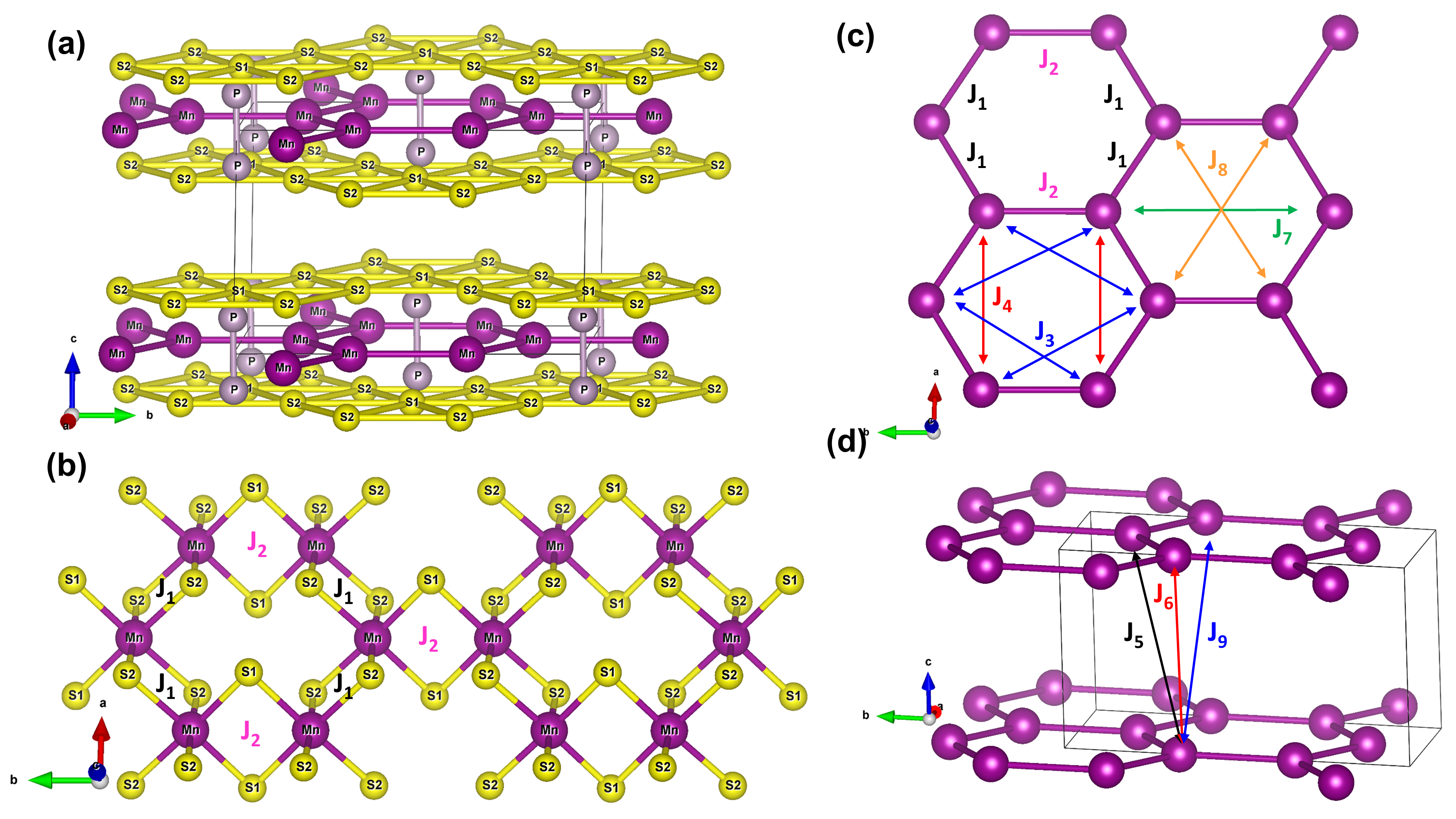}
  \caption{\label{fig:structure} (a) Atomic structure of \mnps.  Mn
    atoms are represented in violet, S in yellow, and P in gray.  (b)
    Detail of the Mn - S bonds in the Mn-S-Mn sandwiches. The $S_1$
    sulfur atoms are associated to the super-exchange path that
    controls the effective magnetic interaction $J_2$. Similarly the
    $S_2$ sulfur atoms are associated to the effective magnetic
    interaction $J_1$. (c) Detail of the six intralayer effective
    exchange interactions considered in this work. The difference
    between $J_1$ and $J_2$, $J_3$ and $J_4$, and $J_7$ and $J_8$ is a
    consequence of the distortion of the honeycomb lattice. (d) Detail
    of the three interlayer effective exchange interactions considered
    in this work. Atomic structures were drawn with VESTA
    \cite{Momma2011}.}
\end{center}
\end{figure*}

This paper is organized as follows. In Sections \ref{sec:exp_details}
we present the crystallographic structure of \mnps\ and the details of
the Raman experiments. In Section \ref{sec:calc_details} we discuss
the methods that we used for the calculation of the effective exchange
interactions and the phonons modes. In Sections \ref{sec:exp_results}
we present and discuss our Raman experiments versus temperature and in
Section \ref{sec:calc_results} the results of our
calculations. Section \ref{sec:discussion} is devoted to the
discussion and conclusion.

\section{Experimental details}
\label{sec:exp_details}

\subsection{\mnps\ structure}
\label{sec:struct_details}

\mnps\ has a monoclinic structure, space group C2/m with lattice
parameters $a$ = 6.077 \AA, $b$ = 10.524 \AA, $c$ = 6.796 \AA, and
$\beta = 107.35\sp{\circ}$ \cite{Ouvrard1985}. The atomic structure is
shown in Figures~\ref{fig:structure}(a) and (b).
There are 4 \mnps\ units in the monoclinic C2/m unit cell, giving a
total of 20 atoms.  The C-centring means that the unit cell is not
primitive because there are two equivalent positions, one at (0, 0 ,0)
and the other at (1/2, 1/2, 0).
A primitive triclinic unit cell containing 10 atoms and having the
minimal volume ${V = 1/2\ a b c \sin \beta}$ can be constructed with,
for example, Bravais lattice vectors
${\vec{a}_1 = (a \hat{x} - b \hat{y}) /2}$,
${\vec{a}_2 = (a \hat{x} + b \hat{y}) /2}$, and
${\vec{a}_3 = c\ (\cos \beta\ \hat{x} + \sin\beta\ \hat{z})}$.

In the paramagnetic phase all the Mn and P atoms are equivalent by
symmetry, while there are two kinds of S atoms. The corresponding
Wyckoff positions of the C2/m cell are 4g, 4i, 4i, and 8j for the Mn,
P, S1, and S2 atoms, respectively.

As can be seen in Fig. \ref{fig:structure}(a), \mnps\ is formed by the
stacking along the $c$ axis of PS-Mn-PS sandwiches. The Mn atoms form
a distorted honeycomb lattice and the S atoms two distorted triangular
lattices. The P atoms can be described as forming P$_2$ dimers
penetrating the Mn hexagons (see Fig. \ref{fig:structure}(a)). Each Mn
atom is octahedrally coordinated to six S atoms which can be
associated to super-exchange paths between the Mn atoms (see
Fig. \ref{fig:structure}(b) and the text below).

\subsection{Raman scattering experiments}
\label{sec:raman_details}

Raman scattering measurements were carried out on single crystals of
manganese phosphorus trisulphide (\mnps) using, as the excitation
source, either a 515 nm line of a continuous wave laser diode or a
514.5 nm line of a continuous wave Ar$\sp{+}$ laser.
The crystals used for experiments were either home-made (see
Ref. \cite{Wildes1998} for more details) or commercially available,
both types of specimens displaying essentially identical spectra. In
accordance to earlier studies \cite{Materials2019}, the Raman
scattering in our samples has been found to be particularly efficient
when using the laser excitation wavelengths around 515 nm.
The collected signal was dispersed through a 0.5 m monochromator and
detected by a CCD camera.

The temperature dependent Raman scattering response was measured at
zero magnetic field with the aid of a continuous flow cryostat mounted
on x-y motorized positioners.
The sample was placed on a cold finger of the cryostat.
The excitation light was focused by means of a 50x long-working
distance objective with a 0.5 numerical aperture producing a spot of
about 1 $\mu$m and the scattering signal was collected via the same
objective.

Magneto-Raman scattering was measured in Faraday and Voigt
configurations using an optical-fiber-based experimental setup. The
sample was mounted on top of an x-y-z piezo-stage kept in gaseous
helium at T= 4.2 K and inserted into a magnet. The excitation light
was coupled to an optical fiber with a core of 5 $\mu$m diameter and
focused on the sample by an aspheric lens (spot diameter around 1
$\mu$m).  The signal was collected by the same lens, injected into a
second optical fiber of 50 $\mu$m diameter. In all experiments, the
excitation power was kept below 100 $\mu$W, to avoid the sample
heating by the laser beam.
Multiple test measurements of the power dependence of Raman scattering
response have not shown any changes in the shape of spectra in the
range of excitation powers up to 500 $\mu$W.

\section{Calculation details}
\label{sec:calc_details}

For the calculations we have used the \textsc{Quantum Espresso}
\cite{Giannozzi2009} code based on density functional theory,
Optimized Norm-Conserving Vanderbilt (ONCV) pseudopotentials and the
PBE functional \cite{Perdew1996b} with a plane-wave and charge-density
cutoff of 90 Ry and 200 Ry, respectively.
An electronic structure calculation with the experimental atomic
parameters \cite{Ouvrard1985} and a PBE functional gives no gap for
the ferromagnetic (FM) order and a gap of 1.30 eV for the
experimentally observed AFM order to be compared with the
experimentally reported gap of 2.96 eV \cite{Gnatchenko2011}.

Including an on-site Hubbard $U_\text{eff}$ on the Mn atoms using the
simplified method developed by Cococcioni and de
Gironcoli \cite{Cococcioni2005} increases the AFM gap up to a
saturation value of 2.60 eV when $U_\text{eff}$ reaches 7 eV.
The same dependence of the electronic gap with the value of
$U_\text{eff}$ has has been reported by other authors, who use either
$U_\text{eff} = 4$ eV \cite{Chittari2016} or 5 eV
\cite{Sivadas2015a,Li2013a}.
A value of $U_\text{eff} = 5$ eV (gap of 2.40 eV) has been used for
the calculations of the effective exchange interactions and
vibrational modes reported below.

The experimental lattice parameters and internal coordinates
\cite{Gnatchenko2011} has been used to calculate the effective
exchange interactions.
As mentioned above, the non primitive 1$\times$1$\times$1 base
centered monoclinic cell contains 20 atoms.
Double monoclinic super-cells (1$\times$1$\times$2,
1$\times$2$\times$1, and 2$\times$1$\times$1) containing 40 atoms and
8 Mn have been used to be able to separate the different exchange
interactions (see below).
We have used a 8$\times$4$\times$2 Monkhorst-Pack \cite{Monkhorst1976}
grid for the first Brillouin zone sampling of the 40 atoms monoclinic
1$\times$1$\times$2 unit cell and adapted equivalent samplings of
8$\times$2$\times$4 and 4$\times$4$\times$4 for the
1$\times$2$\times$1 and 2$\times$1$\times$1 unit cells.

The phonon modes have been calculated in the primitive 10 atoms
triclinic cell with the experimental lattice parameters but fully
relaxed internal coordinates under the constraint of keeping the unit
cell symmetry invariant. After relaxation, the calculation has been
performed using the frozen phonon procedure with a 8$\times$8$\times$8
Monkhorst-Pack \cite{Monkhorst1976} grid for the first Brillouin zone
sampling.

\section{Experimental results}
\label{sec:exp_results}

\begin{figure}[htb!]
\begin{center}
  \includegraphics[width=\columnwidth]{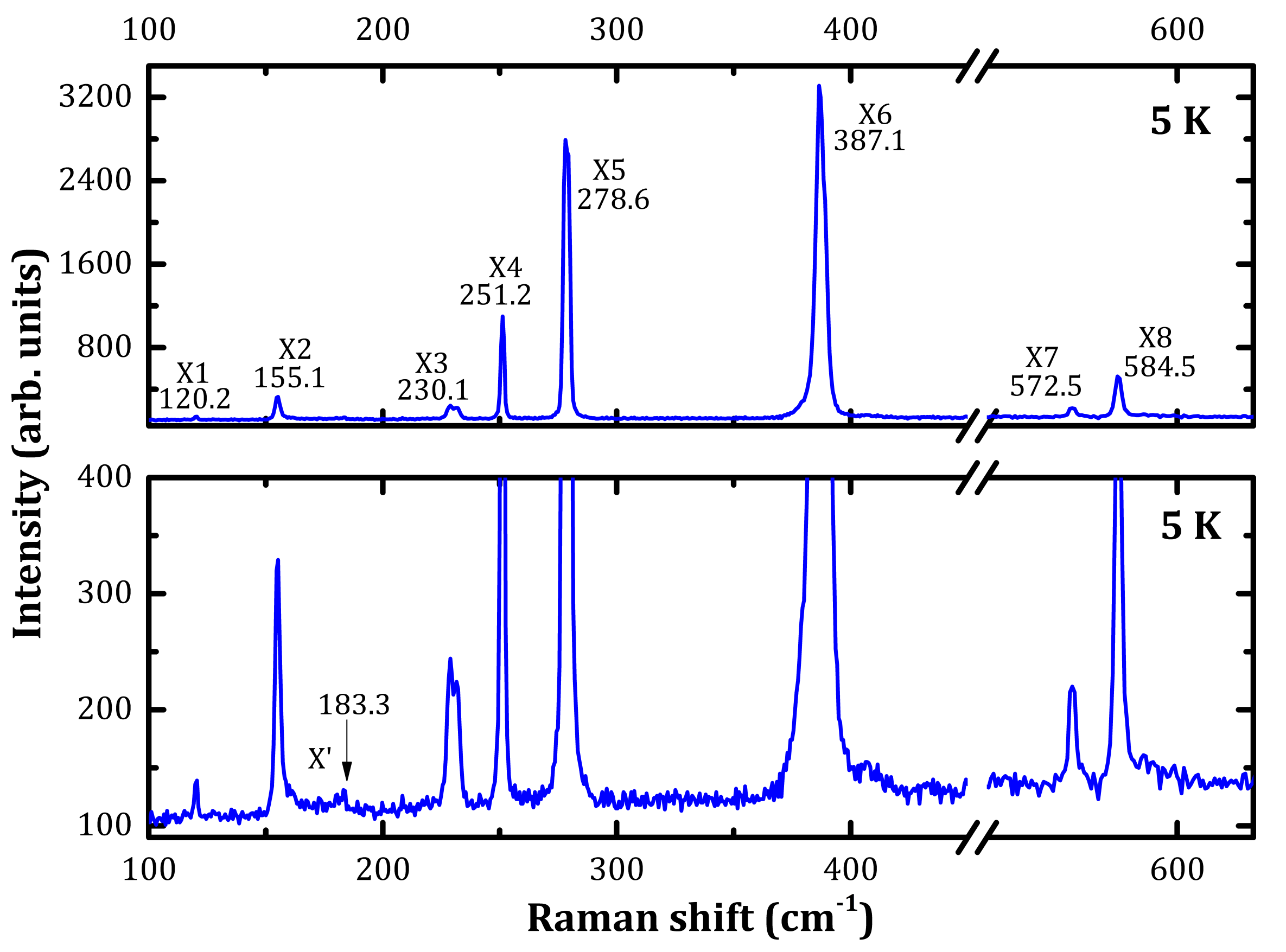}
  \caption{\label{fig:spectra} Characteristic Raman scattering
    spectrum of the \mnps\ crystal, measured under 514 nm laser line
    excitation, at low temperature ($T$ = 5 K) in a wide spectral
    range 100 to 650 \ceme\ from the laser line. The spectrum shown in
    the upper (a) panel is expanded in vertical direction in the
    bottom (b) panel to better visualize the less intense
    Raman-scattering modes. The observed narrow X1 $\dots$ X8 Raman
    scattering peaks are due to phonons whereas a broad X' feature
    with the high energy onset at 183.3 \ceme\ is of different
    origin.}
\end{center}
\end{figure}

The representative Raman scattering spectra of our \mnps\ crystals,
measured at low (5 K) temperature in the range of 100 to 650 \ceme\
from the laser line, are presented in Fig. \ref{fig:spectra}.  A
number of relatively sharp Raman scattering peaks, labeled X1 $\dots$
X8 are observed, in fair agreement with previously reported
observations of the specific phonon modes in \mnps\ crystals
\cite{Mathey1980,Bernasconi1988,DiazMesa2016,Materials2019,Sun2019}.
%
%

\begin{figure*}[htb!]
\begin{center}
  \includegraphics[width=\textwidth]{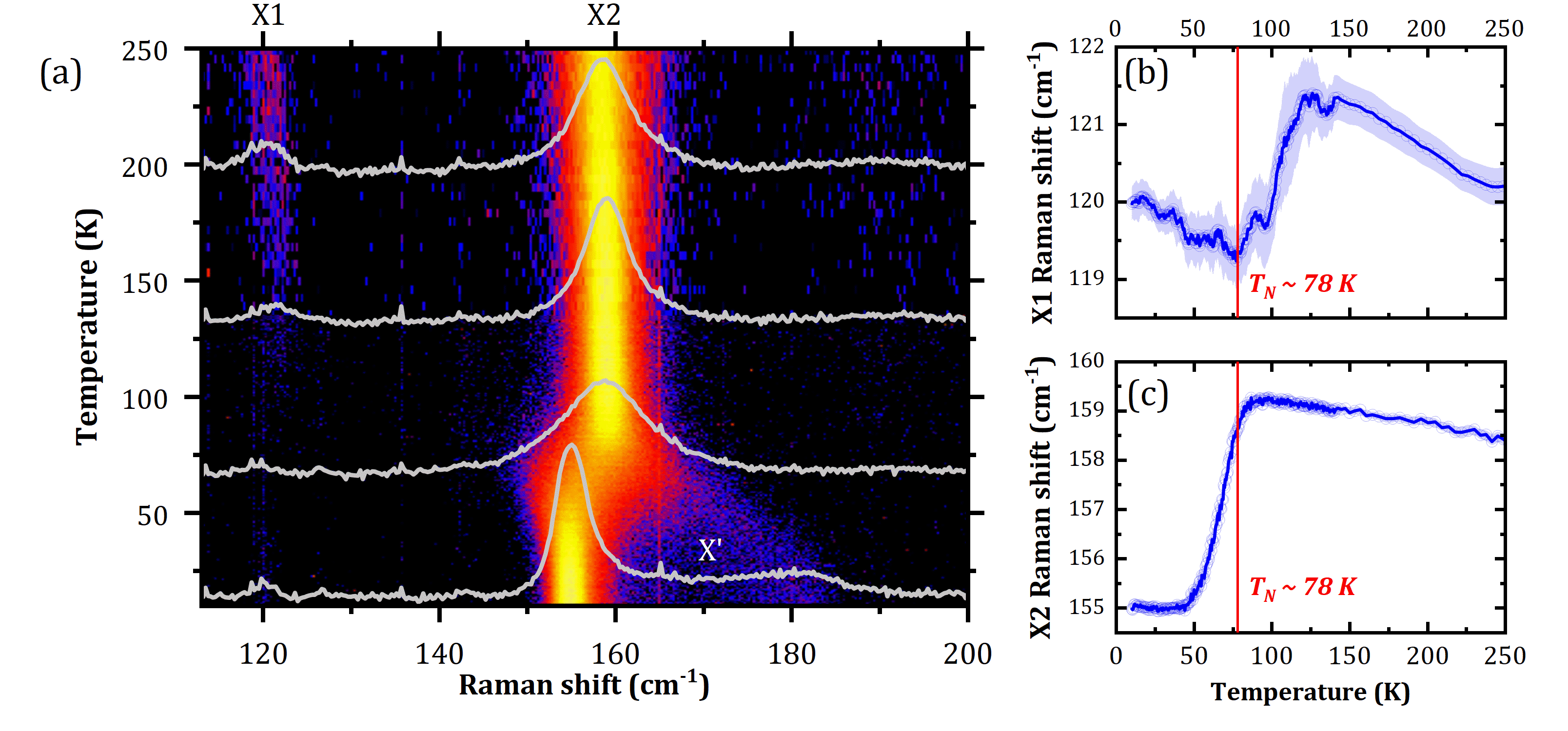}  
  \caption{\label{fig:spectra-vs-T}
    a) False-color map of Raman scattering response (logarithmic scale
    for the intensity) together with a few characteristic spectra
    measured as a function of temperature and focused on the spectral
    region encompassing the X1 and X2 phonon modes which are sensitive
    to the magnetic order in \mnps\ crystals.
    Temperature evolution of central positions of the (a) X1 and (b)
    X2 peaks, as extracted from fitting the shapes of these peaks with
    Lorentzian functions. $T_N$ denotes the critical N\'eel
    temperature for the antiferromagnetic to paramagnetic transition
    in \mnps.}
\end{center}
\end{figure*}

The temperature evolution of all but the X1 and X2 peaks is rather
smooth. Indeed, a conventional red shift and broadening of X3 $\dots$
X8 phonon modes is observed upon increase of temperature (see
Supplementary Material Figures \textbf{S1} and \textbf{S2}). In
contrast, the X1 and X2 Raman scattering peaks appear to be sensitive
to the magnetic ordering in our crystals. These peaks display a
singular behavior around the expected, critical N\'eel temperature,
$T_N$ = 78 K, for the antiferromagnetic to paramagnetic transition in
\mnps\ compound \cite{Wildes2006}. This is illustrated in
Fig.~\ref{fig:spectra-vs-T}.
As can be seen, the center positions of the X1 and X2 Raman scattering
peaks display an abrupt blue shift in close vicinity of $T_N$. In
agreement with previous reports \cite{Materials2019,Sun2019},
%
%
the observed blue shift amounts 5 \ceme\ in the case of the X2 mode
and we report here a similar behavior but with a smaller blue shift of
about 2 \ceme\ for the X1 Raman scattering peak (see
Fig.~\ref{fig:spectra-vs-T}(b) and (c)). More details on temperature
evolution of X1 and X2 peaks as well as of all other X3 $\dots$ X8
peaks are shown in the Supplementary Material's Figures \textbf{S1} to
\textbf{S3}.
Here we note that the X2 peak is relatively intense and well
pronounced in the spectra measured in the whole temperature range,
though it appears to be significantly broadened (by a factor of 4)
around $T_N$ (the maximum in the X2 line-width is observed at
temperature of 65 K, slightly below $T_N$). The X1 peak is relatively
weak. It does not display any obvious broadening around $T_N$ but
displays a relatively abrupt broadening (by a factor of 2) at T
$\simeq$ 110 K.
The intensity of X1 peak initially weakens but significantly increases
when temperature is raised above 100 K.

\begin{figure}[b!]
 \begin{center}
   \includegraphics[width=\columnwidth]{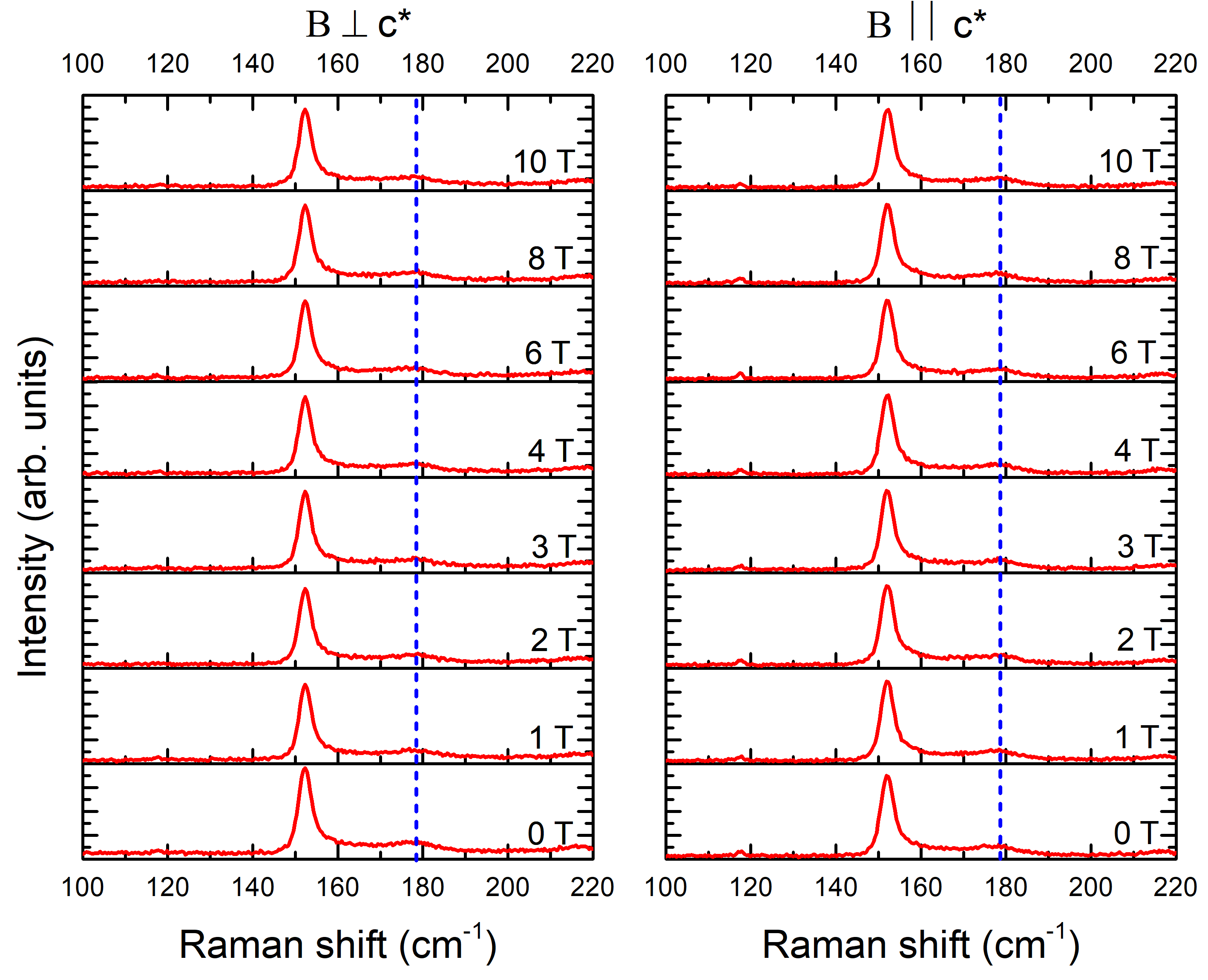}
   \caption{\label{fig:magnon-vs-fieldT} Evolution of the X2
       and X' peaks with an applied magnetic field oriented
       perpendicular (left) and parallel (right) to the reciprocal
       lattice vector $c\sp{*}$. The experiment was performed at 5 K.}
 \end{center}
\end{figure}

\begin{figure}[htb!]
\begin{center}
  \includegraphics[width=\columnwidth]{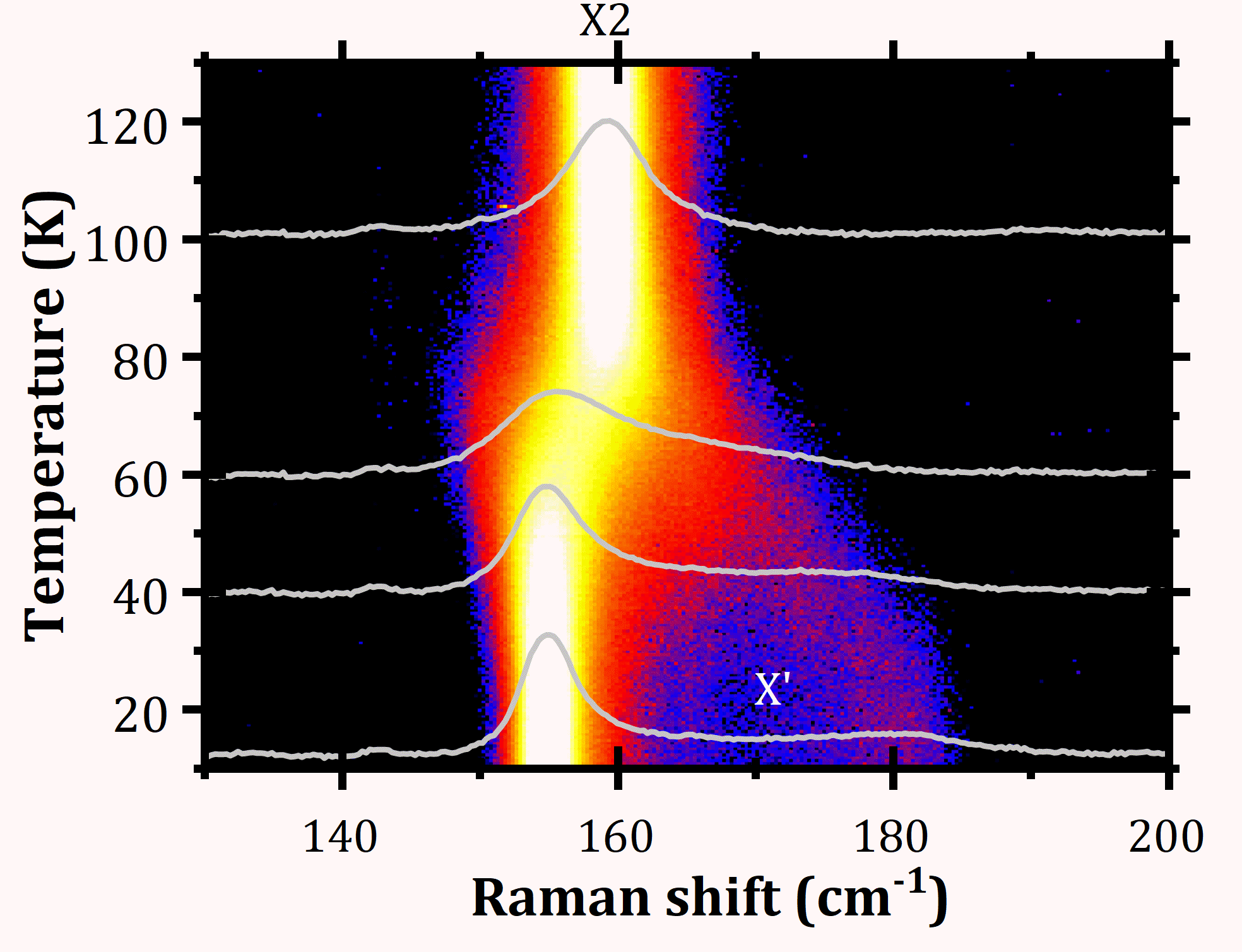}
  \caption{\label{fig:spectra-vs-T-magnon} False-color map of Raman
    scattering response (logarithmic scale for the intensity),
    together with few selected spectra, measured with long acquisition
    times and fine temperature interval of 2 K in the spectral region
    100 to 200 \ceme\ and temperature range from 5 K to 100 K. The
    presented data emphasize a characteristic ''narrowing'' of the X'
    feature with temperature (red shift of its high energy onset),
    which eventually merges with the X2 phonon mode at $T_N$. This
    behavior suggests that the X' feature is due to spin (possibly
    two-magnon) excitations in the antiferromagnetic phase of \mnps.}
\end{center}
\end{figure}

It is interesting to note that, as illustrated in
Fig.~\ref{fig:magnon-vs-fieldT}, both X1 and X2 peaks are not
sensitive to the applied magnetic field.
As already perceived in a recent work \cite{Materials2019}
%
%
the low temperature Raman scattering spectra of \mnps\ crystals show an
additional weak but intriguing feature at energies slightly above the
X2 phonon peak.
This additional feature marked as X' in Fig. \ref{fig:spectra}(b),
might be seen as a continuous-like spectrum with a peak-like,
high-energy onset around 183.3 \ceme\ at low temperatures.
A weak, Raman scattering feature centered at 185 \ceme\ has been
previously observed in \mnps\ compounds studied at room temperature
\cite{Mathey1980}, though it is not observed in our experiments.
Instead, as shown in Fig.~\ref{fig:spectra-vs-T-magnon}, our Raman
scattering studies performed with longer accumulation time and fine
temperature intervals, show that the X' feature exhibits a
characteristic narrowing effect (red shift of the high energy onset)
upon the increase of temperature.
It merges with the X2 phonon peak at temperatures around $T_N$ and it
is not observed in the spectra measured above $T_N$.

The X' feature is likely of a magnetic origin, and the first
  guess would be that it is a due to a coupled phonon (X2)-spin-wave
  (magnon) excitation.
  Such coupling would be favored for $k \approx 0$ spin-wave modes
  whose energy as measured by the spin gap in the magnon dispersion
  relation \cite{Wildes1998,Hicks2019} is only a few \ceme\ ($\sim$
  0.5 meV $\sim$ 4.03 \ceme).
This is in contradiction to the rather large extension ($\sim$ 20
\ceme) of the X’ feature as measured at low temperatures (see
Fig.~\ref{fig:spectra-vs-T-magnon}).
The above hypothesis can also be ruled out by our experiments
performed as a function of the magnetic field (see
Fig.~\ref{fig:magnon-vs-fieldT}) in which we do not observe any
magnetic-field-induced change in the form of the X’ feature.
Very likely, and in accordance with the recent report
\cite{Materials2019}, the X’ feature is due to two-magnon excitations.
Indeed, as deduced from neutron scattering studies \cite{Wildes1998}
the (single) magnon excitations at the Brillouin zone boundary
saturate at energies $\sim $ 11.5 meV $\sim$ 92 \ceme, thus the
two-magnon density of states is expected to peak at around 184 \ceme,
in accordance to the energy position of the onset of the X’ feature
measured at low temperatures.
Notably, the magnon dispersion in the long-wavevector limit, i.e.,
close to the zone-boundary, may persist in the paramagnetic phase
(above $T_N$) but the Raman scattering response due to two-magnon
excitations can be modified at the antiferromagnetic or
paraferromagnetic phase transistion
\cite{Fleury1967,Fleury1969}. Nevertheless, it remains surprising that
our experiments do not show any trace of these two-magnon excitations
at temperatures above $T_N$ \cite{Fleury1967,Fleury1969}.

\section{Calculation results}
\label{sec:calc_results}

\subsection{The magnetic Hamiltonian}
\label{subsec:hamiltonian}

To understand the possible link between the vibrational modes and the
magnetic order in \mnps\ one needs to determine the model magnetic
Hamiltonian of the interacting Mn$\sp{2+}$ ions.

If one considers that these ions are in a 3$d^5$ electronic
configuration corresponding to a high spin $^6S$ (S = 5/2, L = 0)
ground state \cite{Wildes2006}, the effect of the spin-orbit coupling
would be negligible \cite{Joy1992}. The observation of a highly
isotropic magnetic susceptibility above $T_N$ sustains this idea
\cite{Okuda1986,Joy1992,Wildes1994,Wildes1998,Goossens2000b} and
suggests that the effective exchange interactions $J_{ij}$ between the
magnetic centers are isotropic in nature.

Nevertheless, the AFM order below $T_N$ is anisotropic. It was first
reported with the Mn spin moments pointing along the reciprocal
lattice vector $c\sp{*}$, i.e., perpendicular to the honeycomb lattice
in the $ab$ plane
\cite{Kurosawa1983,Okuda1986,Wildes1994,Wildes1998,Wildes2006} but
later on, a more careful analysis of the neutron diffraction data has
shown a small tilt ($\sim 8 \sp{\circ}$) of the Mn$\sp{2+}$ magnetic
moments with respect to $c\sp{*}$ \cite{Ressouche2010}.
Two posible energetic contributions can be proposed to explain the
anisotropic order, the dipolar interactions between the magnetic
moments or a single ion anisotropy arising from a non negligible
spin-orbit coupling.
Theoretical analysis \cite{Pich1994,Pich1995,Pich1995a,Goossens2010}
and recent magnon band measurements \cite{Hicks2019} suggest that the
dipolar interaction would be the leading source of anisotropy
responsible of the observed long-range order. While other experimental
results (electron spin resonance \cite{Okuda1986,Cleary1986}, critical
behavior \cite{Wildes2006,Wildes2007}, and the observation of a non
negligible orbital contribution to the total magnetic moment
\cite{Kurosawa1983,Wildes1994}), support the idea that the single ion
anisotropy is also playing an important role.

In conclusion the full magnetic Hamiltonian should contain a
spin-independent contribution $\hat{H}_0$, a rotational invariant
Heisenberg part $\hat{H}_{\text{Hei}}$, a dipolar term
$\hat{H}_{\text{dip}}$, and a single ion anisotropy contribution
$\hat{H}_{\text{ion}}$ :
\begin{equation}
\label{eq:Hamiltonian}
   \hat{H} = \hat{H}_0            + \hat{H}_{\text{Hei}}
           + \hat{H}_{\text{dip}} + \hat{H}_{\text{ion}} 
\end{equation}
\begin{equation}
\label{eq:Heisenberg}
\hat{H}_{\text{Hei}} =
\sum_{i > j} J_{ij} \; \mathbf{\hat{S}_{i}} \cdot
\mathbf{\hat{S}_{j}}
\end{equation}
\begin{equation}
\label{eq:dip}
\hat{H}_{\text{dip}} = - \frac{\mu_0}{4 \pi}
       \sum_{i > j} \frac{1}{r\sp3_{ij}}
  [ (\mathbf{\hat{m}_i} \cdot \hat{r}_{ij})
    (\mathbf{\hat{m}_j} \cdot \hat{r}_{ij}) -
    \mathbf{\hat{m}_i} \cdot \mathbf{\hat{m}_j} ]
\end{equation}
\begin{equation}
\label{eq:ion}
\hat{H}_{\text{ion}} =
 D\ \sum_{i}\ (\mathbf{\hat{S}_{i}} \cdot \mathbf{\hat{n}})\sp{2}
\end{equation}
where $J_{ij}$ are the magnetic couplings, $\mathbf{\hat{S}_{i}}$ and
$\mathbf{\hat{S}_{j}}$ are the $S=\frac{5}{2}$ spin operators
localized on the Mn atoms at site $i$ and $j$, respectively,
$\mathbf{\hat{m}_i} = g \mu_B\ \mathbf{\hat{S}_{i}}$ and
$\mathbf{\hat{m}_j} = g \mu_B\ \mathbf{\hat{S}_{j}}$ the corresponding
spin only magnetic moments, and $\hat{n}$ the direction of the single
ion easy axis.
Notice that our Heisenberg interaction is defined as
$J\ S_i \cdot S_j$, similar to Le Flem et al \cite{LeFlem1982},
Sivadas et al \cite{Sivadas2015a}, and Chittari et al
\cite{Chittari2016}, while other authors use $-2 J\ S_i \cdot S_j$
\cite{Okuda1986, Joy1992, Wildes1994, Wildes1998, Goossens2000b}.

\subsection{Effective exchange interactions}
\label{subsec:jotas_results}

The full analysis of the magnetic Hamiltonian will be discussed in a
forthcoming paper. To understand the link between the vibrational and
magnetic properties we will only focus here on the rotational
invariant Heisenberg contribution $\hat{H}_{\text{Hei}}$.

We have classified the effective exchange interactions according to
the Mn-Mn distances and we have calculated 9 different interactions up
to interaction distances of 7.647~\AA\ (see Table \ref{table:exchange}
and Figures \ref{fig:structure}(c) and (d)).
\begin{table}[htb!]
\begin{center}
\begin{ruledtabular}
\begin{tabular}{ccccc}
  & d$_\text{Mn-Mn}$ [\AA] & Type  & $J_\textit{i}$ [K] & $\mathcal{J}_i$[K] \\
  \hline
  $J_1$          & 3.500 & intralayer & 12.9 
                 & 19.2 \cite{LeFlem1982}
                   18.0 \cite{Okuda1986}
                   18.2 \cite{Joy1992} \\
  $J_2$          & 3.523 & intralayer & 12.5
                 & 16.2 \cite{Joy1992}
                   17.9 \cite{Wildes1998}
                   18.3 \cite{Sivadas2015a} \\
  \\
  $J_3$          & 6.076 & intralayer & 0.6
                 & \multirow{2}{*}{1.6 \cite{Wildes1998}
                                   0.9 \cite{Sivadas2015a}} \\
  $J_4$          & 6.077 & intralayer & 0.1 \\
  \\
  $J_5$          & 6.791 & interlayer & 0.2
                 & \multirow{2}{*}{-0.04 \cite{Wildes1998}} \\
  $J_6$          & 6.796 & interlayer & 0.5 \\
  \\
  $J_7$          & 7.000 & intralayer & 5.5
                 & \multirow{2}{*}{4.2 \cite{Wildes1998}
                                   5.3 \cite{Sivadas2015a}} \\
  $J_8$          & 7.025 & intralayer & 5.0 \\
  \\
  $J_9$          & 7.648 & interlayer & 0.1 \\
\end{tabular}
\end{ruledtabular}
\caption{Effective exchange interactions.  The nine interactions
  calculated in this work between the Mn atoms obtained using
  density functional theory are listed in the first column.
  The Mn-Mn distances appear in the second column.
  The third column gives the interaction type, either, intra or
  interlayer.
  In the fourth column, the effective interactions are given in units
  of K. A positive value is associated to an antiferromagnetic
  interaction (see text and Eq. (\ref{eq:Heisenberg}) for the
  definition of the exchange interactions used in this work).
  In the last column we show some reported values of the corresponding
  interactions considering the system as having a 2D undistorted
  honeycomb lattice.  The leading exchange interactions are in-plane
  and correspond to the first $\mathcal{J}_1$ ($J_1$, $J_2$) and third
  $\mathcal{J}_3$ ($J_7$, $J_8$) neighbors interactions of the
  undistorted honeycomb lattice.  Interestingly, in this system, the
  in-plane second neighbor interactions $\mathcal{J}_2$ ($J_3$, $J_4$)
  is almost negligible.}
\label{table:exchange}
\end{center}
\end{table}
It is important to note that the distinction made here between the six
(three pairs) of intralayer interactions is a consequence of the small
distortion of the 2D honeycomb lattice.  Indeed, $J_1$ and $J_2$ can
be associated to a single intralayer first neighbors interaction in an
undistorted honeycomb lattice $\mathcal{J}_1$, $J_3$ and $J_4$ to a
single second neighbors interaction $\mathcal{J}_2$, and $J_7$ and
$J_8$ to a third neighbors interaction $\mathcal{J}_3$. Concerning the
interlayer interactions, $J_6$ is the one strictly parallel to the $c$
axis that can be associated to the single interlayer interaction
$\mathcal{J}'$ considered by other authors.

The calculations of the exchange interactions were performed using a
broken-symmetry formalism, i.e., by mapping total energies
corresponding to various collinear spin arrangements within a
supercell onto a Heisenberg
Hamiltonian \cite{Radtke2010a,Saul2011,Saul2013,Saul2014,Radtke2015,Saul2018}.
Different supercells were needed to distinguish the exchange
interactions between two different Mn atoms which are connected by
translation symmetry.
For example, the 40 atoms 2$\times$1$\times$1 unit cell, whose total
energy can be written as :
\begin{eqnarray}
  \nonumber
  E\sp{211} & = & E_0 + \Bigl( \frac{5}{2} \Bigr)\sp2\ 
                  [ a_1\ (J_1 + J_{5}) + a_2\ (J_2 + J_7) \\
  \nonumber
            & + & a_3\ J_3 + a_4 \ (J_4 + J_6) 
              +   a_8\ J_8 + 8\ J_9 ]      
\label{eq:ene211}
\end{eqnarray}
neither allow the calculation of $J_9$ nor the separation of $J_1$ and
$J_5$; $J_2$ and $J_7$; and $J_4$ and $J_6$.
Similar limitations arise with the 40 atoms 1$\times$2$\times$1 and
1$\times$1$\times$2 unit cells, whose total energies are :
\begin{eqnarray}
  \nonumber
  E\sp{121} & = & E_0 +  \Bigl( \frac{5}{2} \Bigr)\sp2\ 
                  [ b_1\ J_1 + b_2\ (J_2 + J_7 + 2 J_8) \\
  \nonumber
            & + & b_3\ J_3 + 8\ J_4 + b_5\ J_5
              +   b_6\ (J_6 + J_9) ] 
\label{eq:ene121}
\end{eqnarray}
and
\begin{eqnarray}
  \nonumber
  E\sp{112} & = & E_0 +  \Bigl( \frac{5}{2} \Bigr)\sp2\ 
                  [ c_1\ ( J_1 + J_5 ) + c_2\ (J_2 + 2 J_8 ) \\
  \nonumber
            & + & c_3\ J_3 + 8\ J_4 + 8\ J_6 + c_7\ J_7 + 8\ J_9]
\label{eq:ene1112}
\end{eqnarray}
The coefficients $a_j$, $b_j$, and $c_j$ depend on the spin
arrangements of the Mn atoms.
As every double supercell contains 8 Mn atoms there are a total of 256
distinct spin configurations. However, on taking crystal and spin
reversal symmetries into account this number can be reduced to 22, 28,
and 25 for the 2$\times$1$\times$1, 1$\times$2$\times$1, and
1$\times$1$\times$2 supercells respectively.

The application of a least-squares minimization procedure to the three
supercells allows us to calculate the nine exchange interactions.
A summary of the calculated values is shown in Table
\ref{table:exchange}.
There are four leading interactions ($J_1$, $J_2$, $J_7$ and $J_8$),
which are all positive (AFM).  They do not lead to frustration and
they are compatible with the experimentally observed AFM order at low
T \cite{Okuda1986}.
As mentioned above, for an undistorted lattice they would correspond
to the intralayer interactions $\mathcal{J}_1 \simeq 12.7$ K,
$\mathcal{J}_2 \simeq 0.3$ K, and $\mathcal{J}_3 \simeq 5.3$ K which
are in good agreement with reported values in the literature.

The distinction between the exchange interactions due to the small
distortion will be important in the analysis of the magnetoelastic
interactions because the distortion splits the energy of some two-fold
degenerate Raman modes.

\subsection{Phonon modes}
\label{subsec:raman_results}

To understand the reasons of the peculiar dependence with temperature
of some Raman modes presented in Section \ref{sec:exp_results}, the 30
phonon modes of \mnps\ have been calculated using the frozen phonon
method in the 10 atoms primitive cell. The irreducible representations
at the $\Gamma$-point are
$\Gamma_{\text{tot}} = 8 A_g + 6 A_u + 7 B_g + 9 B_u$, where only the
8 $A_g$ and 7 $B_g$ modes are Raman Active.

\begin{figure}[htb!]
\begin{center}
   \includegraphics[width=\columnwidth]{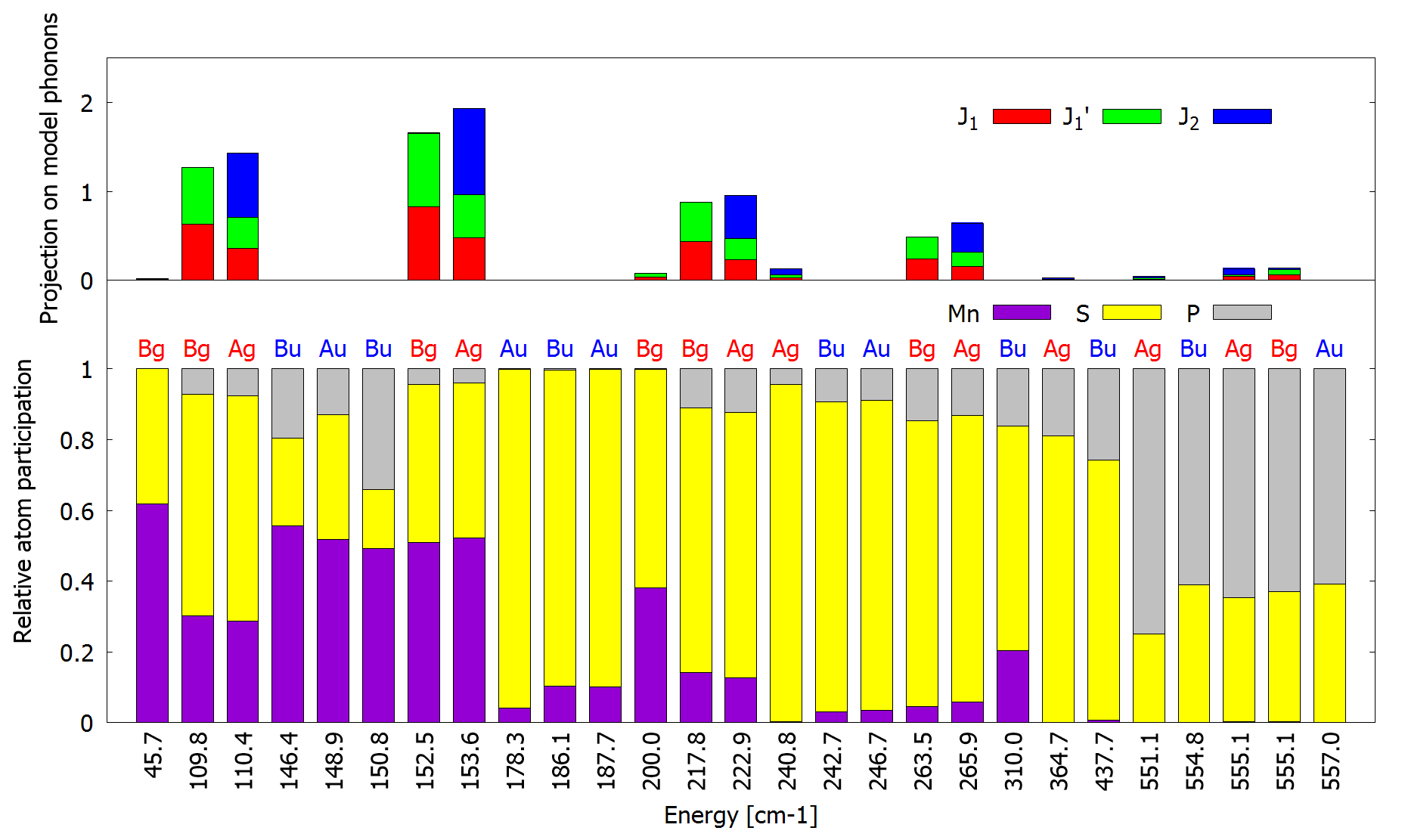}
   \caption{\label{fig:phonons-ldos} 
     Upper panel: projection of the optical phonon modes in the three
     virtual modes shown in Fig.~\textbf{S4} in Supplementary Material. 
     Lower panel: relative atomic participation in the optical phonon
     modes of \mnps\ of the three atom's types. The colors in the bars
     represent the contribution of the two Mn atoms (violet), the six
     S (yellow), and the two P atoms (gray). The symmetry of the modes
     is shown in the upper part of the figure.}
\end{center}
\end{figure}

The calculated 27 optical phonon frequencies that span from 46 to 557
\ceme\ are shown in Fig.~\ref{fig:phonons-ldos}. In the lower panel we
show schematically the energy of the modes, their symmetry and the
relative participation weight of the three atom's types.  For every
mode, the corresponding bars are calculated with the normalized
eigenvectors $\vec{q}$\ of the standard eigenvalue problem:
\begin{equation}
  \mathbf{M}\sp{-1/2}\ \mathbf{H}\ \mathbf{M}\sp{-1/2}\ \vec{q} =
  \omega\sp2\  \vec{q}
\end{equation}
where $\vec{q} = \mathbf{M}\sp{1/2}\ \vec{x}$, $\mathbf{M}$ is the
mass matrix, $\mathbf{H}$ the Hessian of the potential energy, and
$\vec{x}$ the vector with the atomic displacements.
As can be seen from the histogram, the high energy modes are mostly
localized in the P and S atoms while the low energies ones concern
vibrations of the Mn atoms (see Fig.~\ref{fig:phonons-ldos}).

The symmetry and energy of the calculated Raman active modes are also
shown in Table \ref{table:modes}. They compare reasonably well with
several reported values obtained from previous calculations
\cite{Bernasconi1988,Materials2019}. As can be seen, the energies of
some pairs of modes, separated by horizontal lines in Table
\ref{table:modes}, are almost equal because they would be double
degenerate $E_g$ modes in a perfect honeycomb lattice.
Experimentally, it is not always easy to resolve the splitting, even
though the splitting has been sucessfully resolved on feature X3 in
Fig. \ref{fig:spectra}(b) (see also Supplementary Material
Fig.~\textbf{S1}).

\begin{table}[htb!]
\begin{center}
\begin{ruledtabular}
  \begin{tabular}{cccc}
    \multicolumn{2}{c}{Calculations} & \multicolumn{2}{c}{Experiments}  \\
    Symmetry & Energy (\ceme) & Energy (\ceme) & Feature \\
    Bg &      45.7   \\
    \hline
    Bg &      109.8  & \multirow{2}{*}{120.2} &
                                                \multirow{2}{*}{X1} \\  
    Ag &      110.4  \\
    \hline
    Bg &      152.5  & \multirow{2}{*}{155.1} &
                                                \multirow{2}{*}{X2} \\
    Ag &      153.6   \\
    \hline
    Bg &      200.0  \\
    \hline
    Bg &      217.8   &  228.1        & \multirow{2}{*}{X3} \\
    Ag &      222.9   &  231.9  \\
    \hline
    Ag &      240.8  & 251.2 & X4 \\
    \hline
    Bg &     263.5  & \multirow{2}{*}{278.6} &
                                             \multirow{2}{*}{X5} \\
    Ag &     265.9  \\
    \hline
    Ag &     364.7  & 387.1 & X6 \\
    \hline
    Ag &     551.1  & 572.5 & X7 \\
    \hline
    Ag &     555.1  & \multirow{2}{*}{584.5} &
                                             \multirow{2}{*}{X8} \\
    Bg &     555.1  \\
  \end{tabular}
\end{ruledtabular}
\caption{Comparison between the calculated Raman active modes versus
  experimentally observed phonon modes (see Fig.~\ref{fig:spectra} and
  Supplementary Material Fig.~\textbf{S1}).}
\label{table:modes}
\end{center}
\end{table}

The correspondence between the experimental features X1 ... X8 and the
calculated Raman modes is also shown in Table \ref{table:modes}.
There is an overall good agreement between the corresponding energies,
the discrepancy becoming slightly larger for the high energy
modes. For the modes above 200 \ceme, the energy difference increases
with energy giving rise to an overall difference of 5\% for the
majority of the modes while the deviation of the mode around 110
\ceme\ corresponds to a larger relative difference of about 9\%. These
relative deviations of about 5 to 10\% are frequently found when
comparing state-of-the-art DFT calculations with experimental
measurements. It is important to note here that the approximations
used in our calculations have been chosen to have a good overall
description of the magnetic, electronic and structural properties of
the system. A different choice of the DFT input parameters
(pseudopotentials, Hubbard Hamiltonian, functional), may slightly
change the calculated phonon energies and/or the distribution of
deviations between the theoretical and experimental energies of
different modes.
The two calculated modes with irreducible representation Bg around 46
and 200 \ceme\ are not visible in the experimental spectra.

\section{Discussion and summary}
\label{sec:discussion}

\begin{figure}[htb!]
\begin{center}
  \includegraphics[width=\columnwidth]{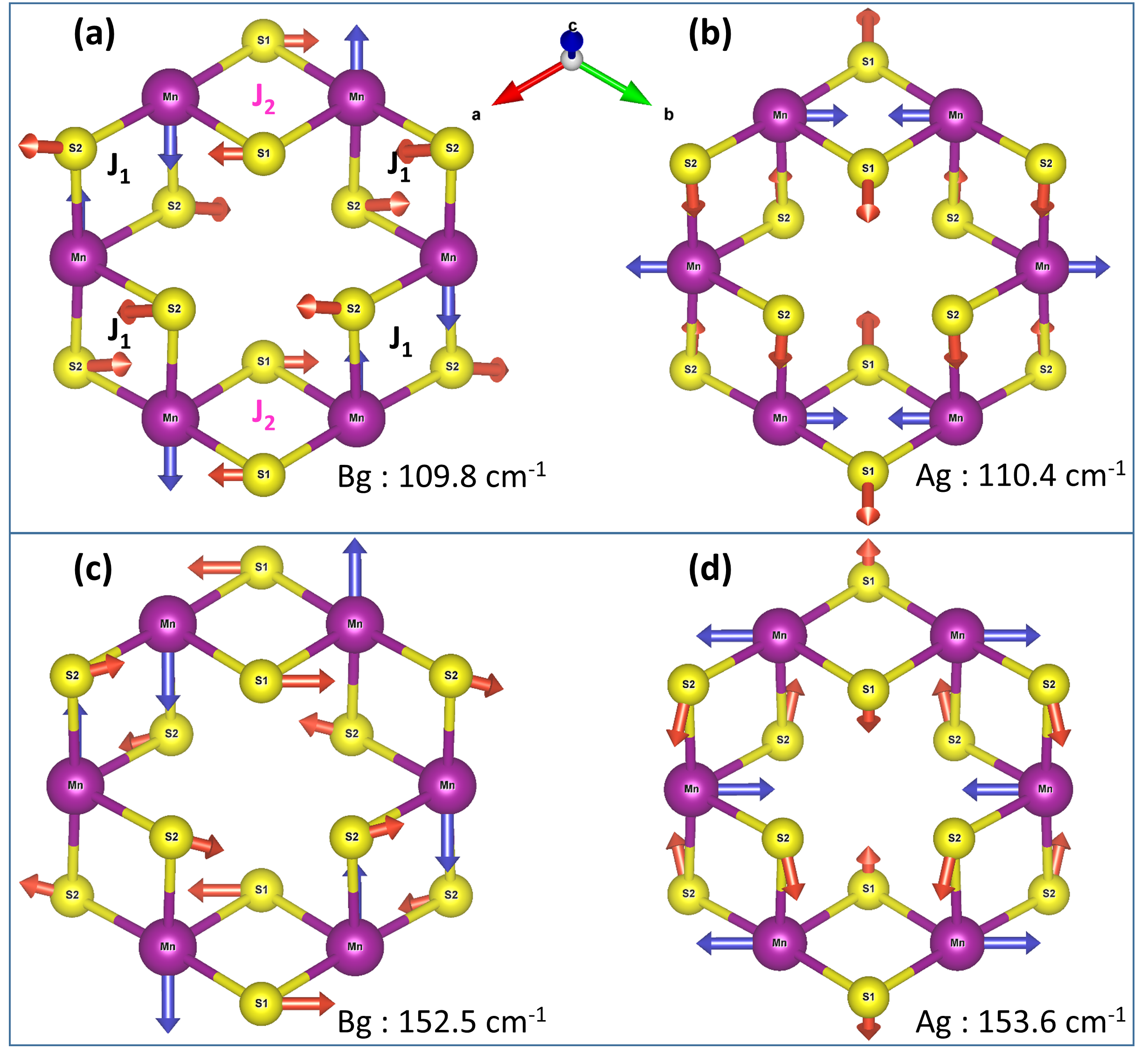}
  \caption{\label{fig:phonons-modes} Schematic representation of the
    displacements of the Mn and S atoms in \mnps\ corresponding to the
    two pairs of Raman active modes associated to the experimental
    peaks X1 and X2.  Phonon mode visualizations were drawn with VESTA
    \cite{Momma2011}.}
\end{center}
\end{figure}

The two Raman modes whose frequencies increase when crossing the
N\'eel temperature at about 120 (X1) and 155 \ceme\ (X2) are
associated to the two pairs of calculated Raman active modes at 109.8
(Bg) and 110.4 (Ag) for X1 and 152.5 (Bg) and 153.6 \ceme\ (Ag) for X2
(see Table \ref{table:modes}). As mentioned above, these two pairs of
almost degenerate modes are split in energy due to the small
distortion of the honeycomb and triangular lattices. For every pair,
the eigenvectors of these almost degenerate modes have very similar
weights in the Mn, S, and P atoms (see lower panel in
Fig.~\ref{fig:phonons-ldos}) and the relative displacements of the Mn
and S atoms for the four modes is shown in
Fig.~\ref{fig:phonons-modes}.

If we consider the six Mn atoms forming the hexagons in the honeycomb
lattice as six consecutive Mn pairs. The effective exchange
interactions of the pairs are either the first or the second neighbors
intralayer interactions $J_1$ and $J_2$ respectively.  These AFM
interactions, which participate to stabilize the N\'eel AFM order at
low temperature, can be associated to superexchange paths between the
Mn atoms through the S atoms (see Fig.~\ref{fig:structure}(b)).

One can imagine that the modification with temperature of these
superexchange paths can be the mechanism at the origin of the
magnetoelastic interactions responsible for the temperature dependence
of the X1 and X2 modes.  Above $T_N$, due the thermal excitations, the
spins localized in some Mn pairs would orient parallel to each
other. As the interaction is AFM, the system would try to decrease the
superexchange angle in order to decrease the interaction. The slight
modification of the angles would explain the energy change of the
Raman modes.

In order to visualize this effect, we have considered three virtual
phonon modes : each one affecting the superexchange angle associated
to two opposite exchange interactions (see Supplementary Information's
Fig.~\textbf{S4}). The upper panel in Fig.~\ref{fig:phonons-ldos}
shows the projections of the 27 optical modes on the three virtual
phonons. The projections are siginificative for four pairs of Raman
active modes and clearly larger for the two pairs of modes asscociated
to peaks X1 and X2.

In conclusion, we have presented a comprehensive picture which largely
accounts for the observed sensitivity of the particular phonon modes
to the magnetic ordering in layered \mnps\ antiferromagnet.  More
specifically, we have identified, both in the experiment and theory,
that there are essentially two, phonon modes of \mnps\, which appear
at 120.2 \ceme\ and 155.1 \ceme\ in the low temperature Raman
scattering spectra, and which show a pronounced change in the vicinity
of the critical N\'eel temperature. The present approach can be
extended to other magnetic materials including their thin layers, in
reference to tracing the magnetic properties with Raman scattering
response due two phonon excitations.

\begin{acknowledgments}
  The work has been supported by the EU Graphene Flagship project,
    the ATOMOPTO project (TEAM programme of the Foundation for Polish
    Science, co-financed by the EU within the ERDFund)
  \end{acknowledgments}

\bibliographystyle{apsrev4-1}

\bibliography{ms}

\end{document}


\title{Supplementary Material: Magneto-elastic interaction in the
  two-dimensional magnetic material MnPS$_3$ studied by first
  principles calculations and Raman experiments }

\author{Diana Vaclavkova}
\affiliation{Laboratoire National des Champs Magn\'etiques Intenses,
  CNRS-UGA-UPS-INSA-EMFL, Grenoble, France}

\author{Alex Delhomme}
\affiliation{Laboratoire National des Champs Magn\'etiques Intenses,
  CNRS-UGA-UPS-INSA-EMFL, Grenoble, France}

\author{Cl\'ement Faugeras} 
\affiliation{Laboratoire National des Champs Magn\'etiques Intenses,
  CNRS-UGA-UPS-INSA-EMFL, Grenoble, France}

\author{Marek Potemski}
\affiliation{Laboratoire National des Champs Magn\'etiques Intenses,
  CNRS-UGA-UPS-INSA-EMFL, Grenoble, France}
\affiliation{Institute of Experimental Physics, Faculty of Physics,
  University of Warsaw, 02-093 Warsaw, Poland}

\author{Aleksander Bogucki}
\affiliation{Institute of Experimental Physics, Faculty of Physics,
  University of Warsaw, 02-093 Warsaw, Poland}

\author{Jan Suffczy\'nski}
\affiliation{Institute of Experimental Physics, Faculty of Physics,
  University of Warsaw, 02-093 Warsaw, Poland}

\author{Piotr Kossacki}
\affiliation{Institute of Experimental Physics, Faculty of Physics,
  University of Warsaw, 02-093 Warsaw, Poland}

\author{Andrew R. Wildes}
\affiliation{Institut Laue-Langevin, Grenoble, France}

\author{Beno\^it Gr\'emaud}
\affiliation{Aix Marseille Univerist\'e, Universit\'e de
  Toulon, CNRS, CPT, Marseille, France}
\affiliation{MajuLab, CNRS-UCA-SU-NUS-NTU
  International Joint Research Unit, 117542 Singapore}
\affiliation{Centre for Quantum Technologies, National University of
  Singapore, 2 Science Drive 3, 117542 Singapore}

\author{Andr\'es Sa\'ul}
\affiliation{Aix-Marseille Universit\'e, Centre Interdisciplinaire de
Nanoscience de Marseille-CNRS (UMR 7325), Marseille, France}
\email{saul@cinam.univ-mrs.fr}

\maketitle

\begin{figure*}[htb!]
  \includegraphics[width=0.8\textwidth]{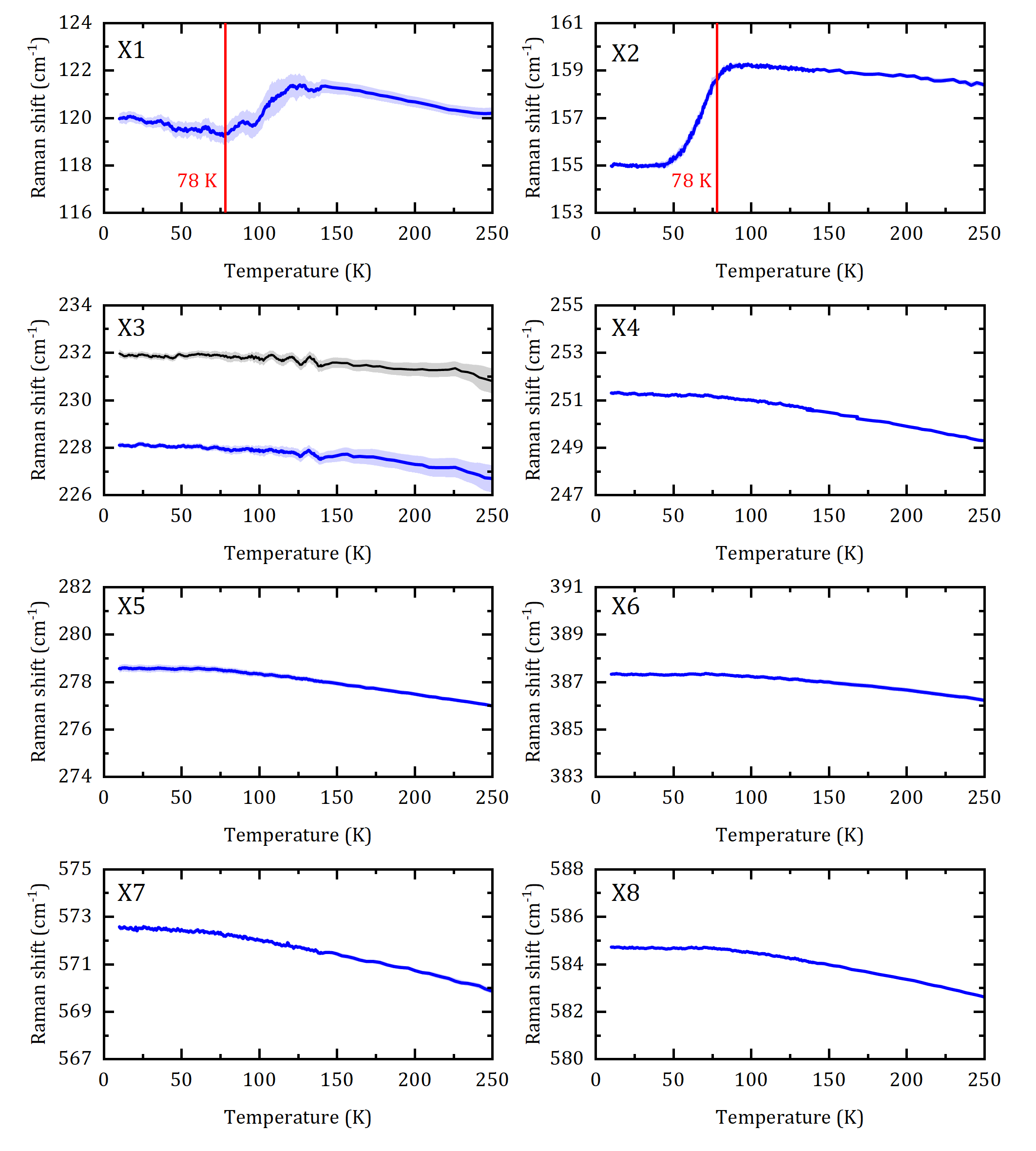}
  \caption{Temperature dependence of the central positions of
    phonon-related, X1, .. X8, Raman scattering lines in \mnps. }
\label{fig:ext1}
\end{figure*}
\begin{figure*}[htb!]
  \includegraphics[width=0.8\textwidth]{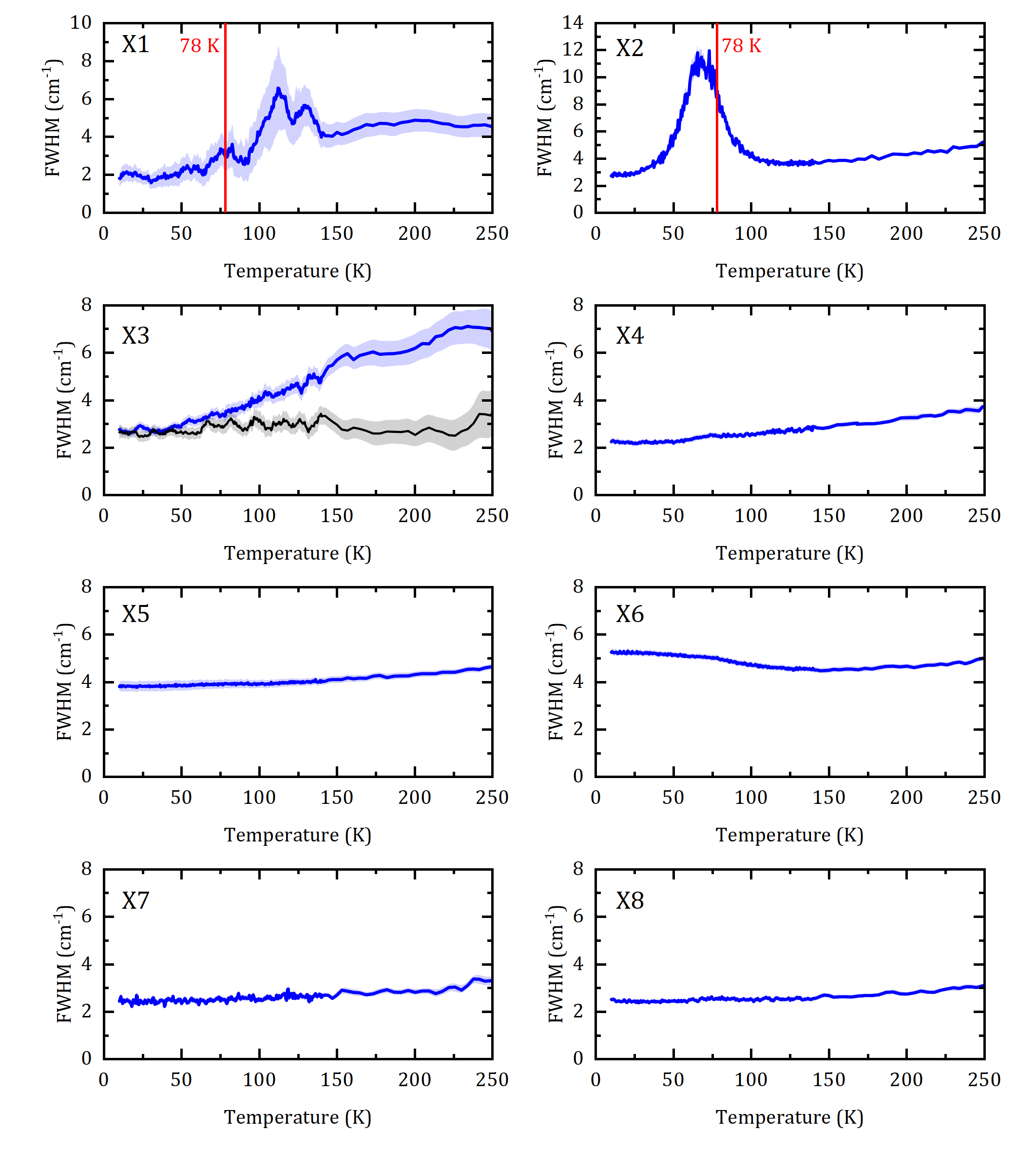}
  \caption{Evolution with temperature of the full width at half
    maximum FWHM of phonon-related, X1, ..X8, Raman scattering lines
    in \mnps\ crystals.}
\label{fig:ext2}
\end{figure*}
\begin{figure*}[htb!]
 \includegraphics[width=0.8\textwidth]{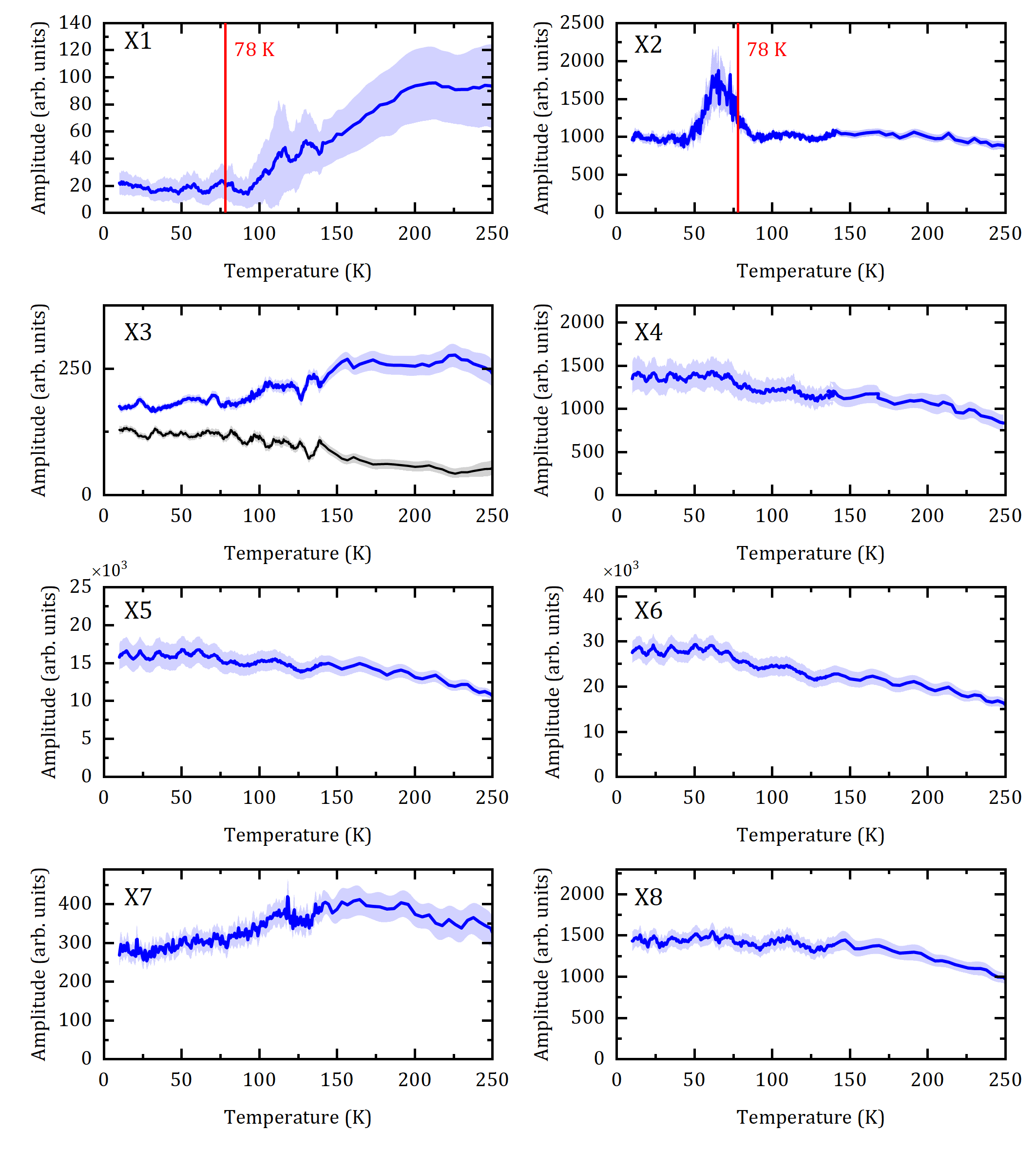}
  \caption{Temperature evolution of the amplitude of phonon-related,
    X1, ..X8, Raman scattering lines in \mnps\ crystals.}
\label{fig:ext3}
\end{figure*}

Additional results concerning the analysis of the Raman scattering
spectra of \mnps, measured as a function of temperature are shown
below.
All Raman scattering lines, X1 ... X8, (for notation, see the main
text) which are visible in our spectra and which have been attributed
to phonon modes has been fit to Lorentzian functions and the central
position, linewidth, and amplitude of these lines have been
extracted. The resulting temperature evolution of these three
parameters is illustrated in Figures \ref{fig:ext1}, \ref{fig:ext2},
and \ref{fig:ext3} for each X1 ... X8 resonance.

\begin{figure*}[htb!]
 \includegraphics[width=0.8\textwidth]{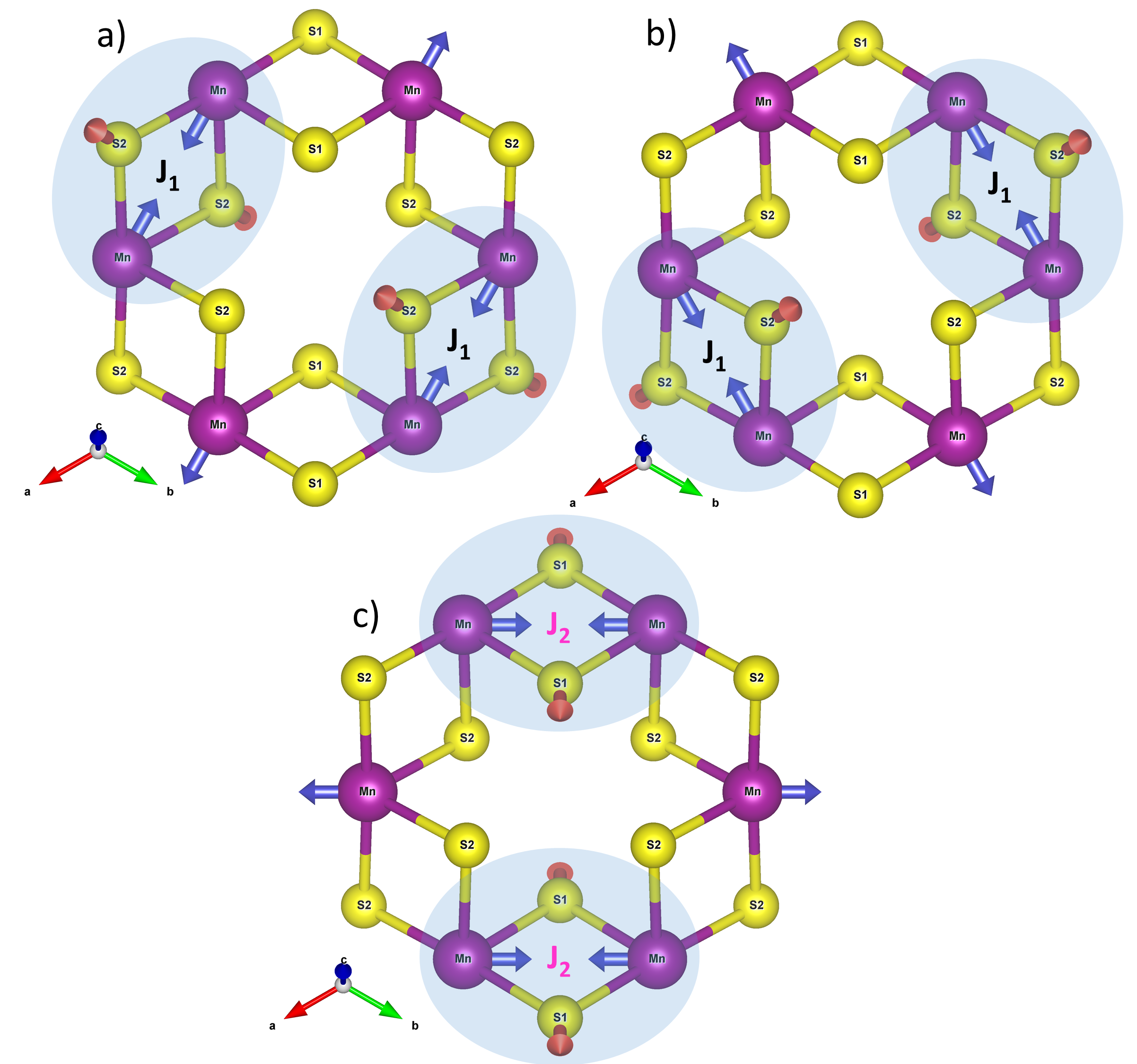}
 \caption{Virtual atom displacements that would change the
   corresponding superexchange angles associated to the effective
   exchange interactions. Phonon mode visualizations were drawn with
   VESTA \cite{Momma2011}.}
\label{fig:virtual}
\end{figure*}

Figure \ref{fig:virtual} show the schematic atom displacement, i.e.,
phonon modes, that modifies the superexchange angles associated with
the effective exchange interactions $J_1$ and $J_2$.

\bibliographystyle{apsrev4-1}

\bibliography{supplementary}